\documentstyle[preprint,tighten,titlepage,eqsecnum,aps,epsf]{revtex}

\def\sgn{\mathop{\rm sgn}\nolimits}
\begin{document}
\preprint{KFKI-1994-24/A}
\title{Are There  Top Quarks in Superdense Hybrid Stars?
}
\author{M. Priszny\'ak, B. Luk\'acs  and  P. L\'evai
       }
\address{
Institute of Nuclear and Particle Physics, Theor. Dep.\\
Budapest, H-1525 P.O.B. 49  \\
E-mail: [surname]@rmki.kfki.hu
}
\date{\today}
\maketitle

\pagenumbering{alph}
\thispagestyle{empty}
\begin{abstract}

 It is examined whether primordial top quarks from the Big Bang
 could survive under natural conditions till present days.
 Hybrid
 quark-neutron stars, with
 dense enough core to maintain the existence of
top quarks, is discussed. In this paper only final states
are investigated.  Zero temperature quark-lepton matter is modelled as
 as a gas of non-interacting particles.
 Local charge neutrality has been assumed
 in order to avoid volume interactions which would not make
 thermodynamical treatment possible.

\end{abstract}
\pacs{  }

\narrowtext

\pagestyle{plain}
\pagenumbering{arabic}
\section{Introduction}
\label{sec:intro}

With the claimed  discovery of the top quark at FERMILAB \cite{FERMILAB}
the quark family has become complete and masses of all quarks are known.
Now we are in the position to
answer  the question
of occurrance of heavy quarks, including the top, in Nature.
Such a scenario may belong to cosmology, more exactly the
physics of Big Bang
or to astrophysics, namely superdense quark stars.
They may have dense enough cores to support the existence of
heavy quarks and our aim is to investigate the necessary conditions for
the existence and stability of these stellar objects.

 According to the standard theories of stellar evolution,  the final
 state of massive stars can be compact, dense stellar objects,
neutron stars, with radius of
 $10$ km and mass of $0.1-2 \; M_{\odot}$. Since the redshift
parameter
is about 0.2 on the surface of these objects one must use general
relativity. For a non-rotating, static
stellar object the Tolman-Oppenheimer-Volkoff\cite{TOV,tolman} equation
 expresses the condition of hydrostatic
equilibrium and connects nuclear physics, through the equation of
state
(EOS), with astrophysics, through  gravitation.
Since equilibrium configurations exist for {\it any} central
densities,
 actually even for infinite central density as well, a quark core
 may support hadron matter in neutron stars.  However stability
against oscillations still must be checked.

It is a generally held belief in the physics community that very dense
nuclear matter ($\rho \geq 5 - 6 \ \rho_{nucl}$) actually appears
in the form of quark-gluon plasma (QGP), i.e. the physical degrees of
freedom are no longer hadron particles but quarks and gluons.

 We assume that locally the QGP matter is electrically neutral.
 Without this assumption we would have a volume
self-interaction, so a non-extensive energy
variable for our system and no thermodynamic approach could be
applied.
Thenceforth one could not tell either what different phases exist
inside the
star because to obtain phase boundaries one needs thermodynamics.

 In Section \ref{sec:primordial} we discuss whether primordial top
quarks could  survive from the Big Bang.
 Section \ref{sec:confinement} is about the confinement transition
 of the Universe. Section \ref{sec:hybrid} deals with hybrid
stars.
 Section \ref{sec:stabil} addresses the question of stability of
 these stellar objects.

We note that this paper is only a  preliminary report of an on-going
research work. At some points serious simplifying assumptions were
used to make the calculations possible. Our main goal was to find
the top peak on the mass-density plot if it exists at all.
A future paper will rediscuss some simplifying assumptions.

\section{ Primordial top quarks from Big Bang?}
\label{sec:primordial}
  Top quarks must have
been
 present in the very early Universe in two ways. Today's
 observations suggest a positive total baryon or quark number in
 our Universe. At
 temperatures higher than $\approx (m_t-m_c)c^2$ a substantial
part
 of this total quark number must have been in $t$ quarks. True
enough, we do not know
 when the present baryon number was created. Besides an
 originally matter-antimatter-asymmetric Universe, which is a
 not too attractive possibility, at least three different ideas
have
 been seriously discussed. Namely, net baryon number may have
 been generated
 \begin{itemize}
 \item  in a non-equilibrium situation just after the
        GUT symmetry breaking, i.e. at cca. $10^{-35}$ s and
$10^{14}$ GeV
        temperature \cite{bar83}
 \item  during the electroweak transition (
        $t \approx 10^{-10}$ s, $T \approx 1000$ GeV) via
anomalous
        processes through e.g.  sphalerons \cite{klink84}
 \item  from  lepton excess produced in an assumed SUSY transition,
        by anomalous processes before the electroweak
        transition\cite{bar83}.
\end{itemize}
  Therefore the terminus ante quem for $B > 0$ seems to be 100
picosec.  However,
  $t$ quarks were present even without a quark excess, in the form of
  $t\bar{t}$ pairs.  Such
  pairs are present in substantial quantity at $T \geq m_t c^2$.
Both
  conditions hold practically at the same temperature, $T \geq
175$ GeV.
  To estimate characteristic times, let us take a
radiation-dominated $k=0$ Friedmann
  universe,
\begin{eqnarray}
    ds^2 =  dt^2 -  R^2 (t) \{ dx^2 +  x^2d^2\Omega \}
\label{eq:frid} \\
   \varepsilon = 3P = (N\pi^2 /30) T^4 \label{eq:erad}
\end{eqnarray}
 where $N$ is the effective number of helicity degrees of freedom, $1$
 for each light boson and $7/8$ for each light fermion pair. In the GUT
 symmetric phase $ N \approx 162$, and just before the breakdown of
 electroweak symmetry $N \approx 10$.  The solution of
 Eqs.(\ref{eq:frid}-\ref{eq:erad}) is
\begin{equation}
 T = \sqrt[4]{ {3 \over{32\pi}} {30 \over{N\pi^2}}} \sqrt{\hbar
{E_{Pl}/t}}
\end{equation}
 where
 the $E_{Pl}$ Planck energy is $1.22 \; 10^{19}$ GeV.
 Hence $T \approx 100$ GeV at $t \approx 100$ picosec.
 Therefore $t\bar{t}$ pairs vanished at $100$ picosec, and excess tops,
 if already existed then, started to decay into charm and up
 quarks. The decay
 process is driven by weak interaction, therefore at this energy
 difference the lifetime is definitely shorter than $100$ picosec.
 Consequently top quarks practically vanished from the Universe
 not later than 100 picosec = $10^{-10}$ s after the Beginning.

\section{ Top quarks in the confinement transition}
\label{sec:confinement}

Bottom quarks start to annihilate and decay at $T \approx 4.5$ GeV,
which corresponds to cca. $10$ nanosec. For a review of the early
history of the Universe see e.g. \cite{kampluk}. Decay times are
much shorter,
so they vanish practically instantaneously. The same happens
with the $c$ quarks at 100 nanosec. The decay/annihilation begins
for $s$ quarks, too, at several microsecs, but that process is
interrupted at $7$ microsecs by a first order confinement transition
\cite{kampf84}.  In this transition 3-quark groups form baryons and
quark pairs form mesons.  The fine details may depend on unevaluated
details of nonperturbative QCD; anyway, serious density
differences can be expected during the transition which lasts
between cca.  $7$ and $12$ microsecs. One possible result is the
formation
of dense quark blobs.  The later fate of these dense blobs is
under debates; some authors claim that the so-called symmetric
quark matter (equal weights of u, d and s) may be stable
against hadron formation\cite{farhi}. But the fate cannot be calculated
without knowing the characteristic sizes of the blobs.  The
transition starts from nucleation cores of the new phases. The
evolution is uncorrelated at distances greater than $ct$ , therefore at
$7$ microsecs the radius of the emerging blobs will be in the
range
of some kilometers.  Consequently the blobs will resemble neutron
stars of the order of solar mass.  As it is well known, a neutron
star is a stationary object sustained in its own gravity, and density
inhomogenities inside can be large. Strictly speaking the central
density of a neutron star has no upper limit; in all
calculations equilibrium configurations exist for even infinite
central density\cite{harr}. (Stability against radial oscillations is
another matter and will be studied in due course.) So, the
phase transition may regenerate the heavy quarks
vanishing earlier. (Similarly can the gravitational collapse
do so, resulting in neutron or quark stars at much later times.)
The
calculation of the equilibrium states
needs full general relativity and
the equation
of state of relativistic QCD plasma.

\section{Top quarks in hybrid stars}
\label{sec:hybrid}

\subsection{The Tolman-Oppenheimer-Volkoff equation of general
relativity}
In general relativity theory there always exists a locally free-falling
frame in which the laws of special relativity govern physics. For a
uniform fluid at rest in this frame the $T_{\mu \nu}$ energy-momentum
tensor defines what we mean by pressure and energy density. Once $T_{\mu \nu}$
is known in a covariant form in this frame it serves as a source term
in the Einstein equation of GTR.

Let us restrict ourselves to static spherical objects.  Then we have
$4$ Killing vectors in the space-time, of which $3$ are space-like, act
on $2$ dimensional transitivity surface and commute as $SO(3)$;
the
fourth one is timelike and commutes trivially with all of the previous
three\cite{misner}.  Then, up to coordinate transformations, the most
general possible line element is of form
\begin{equation}
ds^2 = \exp{(2\nu(r) )} dt^2 - \exp{(2\lambda (r))} dr^2 -r^2
d\Omega \end{equation}
Now, the curvature is governed via the Einstein equation
\begin{equation}
R_{ik} - {1\over 2} g_{ik} R = -{8\pi G } T_{ik}
\label{eq:einstein}
\end{equation}
where $R_{ik}$  is the Ricci tensor and $T_{ik}$ is the energy-momentum
tensor. Henceforth $c=1$. For a fluid the latter one has the form
$T_{ik} = \varepsilon
 u_i u_k - P (g_{ik} - u_i u_k )$.  Here $\varepsilon$ is the energy
density and $P$ is the pressure. Both  can be calculated from the
equation of state.  The velocity field of the fluid is in our case
purely temporal, and of unit length.
\begin{equation}
u_i = (1,0,0,0)  \label{eq:vel}
\end{equation}

For the present symmetries eq. (\ref{eq:einstein}) possesses three
nontrivial and algebraically independent components, say the $00$, $11$ and
$22$ ones. However the velocity field (\ref{eq:vel}) is a solution of the
equation of motion, following from the automatically vanishing divergence
of the left hand side of eq. (\ref{eq:einstein}). So now two components
will suffice. We can write them as follows:
\begin{eqnarray}
8 \pi G T^1_{\; 1} &= 8 \pi G P &= - e^{-\lambda} ( \frac{\nu}{r} +
\frac{1}{r^2} ) +
    \frac{1}{r^2} \\
8 \pi G T^2_{\; 2} &= 8 \pi G P &= - \frac{1}{2} e^{-\lambda}
( \nu\prime\prime + \frac{1}{2} \nu\prime^2 +
 \frac{\nu\prime - \lambda\prime}{r} +
    \frac{\nu\prime \lambda\prime}{2} )
\end{eqnarray}
Now let us write, simply as a {\it definition}:
\begin{equation}
\exp{(-2\lambda(r)) } \equiv 1 - 2 m (r) /r
\end{equation}
Then an integro-differential equation is obtained for $m(r)$, which we
write here as
\begin{eqnarray}
{{dP} \over{dr}} &= -G \frac{ (\varepsilon+P)(m(r)+4\pi P r^3)}
{ r (r-2 m(r))}
\label{eq:TOV} \\
m(r) &= 4 \pi \int_0^r\nolimits \varepsilon ({\tilde r})
 {\tilde r}^2 \; d{\tilde r}
\label{eq:MassCont} \end{eqnarray}
and there remains a quadrature for the function
$\nu{(r)}$, which is trivial and will be ignored here. In the last
equation we exploited a regularity relation $\lim_{r \rightarrow 0}
m(r) = 0$.  Eq. (\ref{eq:TOV}) is the Tolman-Oppenheimer-Volkoff
(TOV) equation.  If the equation of state is fixed as a
function $ P = P(\varepsilon)$ or in any equivalent parametric
form, then
to any central density $\varepsilon (0) = \varepsilon_0$ a unique
solution can be obtained.

  \subsection{Phase transition}

    In our environment the low density state of quarks is
    hadronic matter, i.e. a mixture of tightly bound $qqq$ and
   $q\bar{q}$ clusters. It is not quite sure that this is the
lowest energy
        state indeed; there are speculations \cite{farhi} that the
"symmetric strange matter", i.e. an
        equal weight mixture of $u$, $d$ and $s$ in plasma state,
may be
        energetically preferred against neutron matter, and then the fact
        that nuclear matter remains in hadrons could be explained in
        terms of the macroscopic $S$ charge of such a state. Then
there may
 be a high potential barrier between the local $S=0$ minimum of
nuclear
        matter and the global minimum of $S=B/3$.
  The final answer is not yet known because the calculations are rather
   difficult for dilute quark plasma with QCD being there
nonperturbative.
   Now, observe that the "potential barrier" argumentation may still
    be valid in situations where hadronic matter transforms into quarks
    in a compression process. For the primordial quarks forming a
quark  star between 7 and 12 microsecs the argument is clearly
irrelevant.
        Namely, at that time $s$ quarks still were abundant;
        star-sized blobs of the symmetric ($uds$) plasma can be
formed by using
        up the existing $s$ quarks. (This is roughly the suspected
scenario
   of "strange nugget" formation \cite{witten}.)
  Therefore it is better not to make preconceptions about the
  low-density crust of the stars under investigation. However
  the conditions for phase transition are clear. Hadrons
$h_{\alpha}$ are
   built up from quarks $q_i$ with stochiometric numbers
$c_{\alpha}^i $:
\begin{equation}
    \sum_r c_{\alpha}^r q_r \rightarrow h_{\alpha}
\end{equation}
Then phase equilibrium holds if
\begin{eqnarray}
  \sum_r  c_{\alpha}^r \mu_r &= \mu_{\alpha} \label{eq:equil1} \\
          p_{QGP} &= p_H  \label{eq:equil2}
\end{eqnarray}
    Now, we start from inside where the matter is a quark plasma
    with no doubts. If, going outward, somewhere Eqs.
(\ref{eq:equil1}-\ref{eq:equil2}) holds
 then there a phase transition happens and the TOV equation is to be
  continued with the equation of state of the new phase.

   Since now there are two different quark chemical potentials
  (see Subsection \ref{sec:EOS}) the actual phase structure may be
quite complicated (for a
   discussion see \cite{levluzi}.) However it seems that, if
hadronic phase
        exists, it is dominated by neutrons; e.g. Ref.
\cite{harr} got $7/8$ part $n$
        vs. $1/8$ part $p$. Then for a first approximation the
phase transition is
        \[  3 q \rightarrow n \]
where one quark is $u$, $c$ or $t$, two are $d$, $s$ or $b$; and
Conds. (\ref{eq:equil1}-\ref{eq:equil2})   read as
\begin{eqnarray}
          \mu_u + 2 \mu_d \approx \mu_n \\
          p(q) \approx p(n)
\end{eqnarray}
  This will be our condition to switch to nuclear equation of state.

\subsection{The stability of configurations}
\label{sec:stabil}

 An equilibrium configuration may still be unstable against
 oscillations. The sufficient and necessary conditions of stability
 against oscillations can be found in the B Appendix of
Ref. \cite{harr}. In
   that work a series of theorems has been proven for those who
are
   interested only in the question of the stability  and not in
oscillation frequencies.

    The oscillation frequency squares form a series of disjoint values
         $\Omega^2_i = \Omega^2_i(\varepsilon_0)$
where $\varepsilon_0$ is the central energy density of the
configuration.
A mode becomes unstable where its $\Omega^2_i$ becomes zero from
above.
         A mode can
change its stability
        only in points where $M(\varepsilon_0)$ has an extremum.
Ignoring possible inflexion points, one gets the scheme that
(in parentheses the shorthand terms referring to directions of
turns on an $R(M)$ diagram)
for maximum
\begin{itemize}
\item with $R(\varepsilon_0)$ increasing (left turn on the right
hand side) : the last unstable
mode becomes stable;
 \item    with $R(\varepsilon_0)$ decreasing (right turn on the
right hand side) : the first stable
mode becomes unstable;
\end{itemize}
for minimum
\begin{itemize}
\item  with $R(\varepsilon_0)$ increasing (right turn on the left
hand side) : the first stable mode
becomes unstable;
\item     with $R(\varepsilon_0)$ decreasing (left turn on the left
hand side) : the last unstable
mode becomes stable; \end{itemize}
   (See also Ref. \cite{bluk}.) Since until now all calculations
for the neutron star
peak gave stability on the slope upwards, stability of all peaks can be
read off from
the  $M(\mu)$
and $R(\mu)$
curves  if they are sufficiently reliable in details.

\subsection{The equation of state
           }
\label{sec:EOS}

At first we assume that the core of the superdense quark star can
be described
as non-interacting Fermi gas of quarks and leptons at zero
temperature. Having no internal energy production, these stellar
objects will quickly cool to near absolute zero\cite{kihul}.

Thus the degrees of freedom taken into account are fermionic.
 The r\^ole of other bosonic degrees of freedom  (e.g.
quasi-particles made up of interacting fermions) will be
discussed in a subsequent work. At zero temperature we need not
deal
 with antiparticles either.  The effect of
nonperturbative interaction among quarks
will be simulated by introducing the bag constant, $B$, which is
important
at lower densities to determine the interface between the quark core and the
hadronic surface layer. This value of $B = 260$ MeV is derived
from phenomenological high energy physics and may not be quite
suitable for matter at these densities but this is the best value available
now.

Since on cosmic time scale the weak interactions will assure chemical
equilibrium, one can introduce chemical potentials for all particle species
and linear equations will hold among them.
The chemical equilibrium is established by the following one-W exchange
processes:
\begin{eqnarray}
 s + u &\leftrightarrow d + u  \nonumber \\
 c + s &\leftrightarrow u + s  \nonumber \\
 b + c &\leftrightarrow s + c  \nonumber \\
 t + b &\leftrightarrow c + b
\end{eqnarray}
These equations imply the following relations for the chemical potentials of
the differently charged members of the quark dublets:
\begin{eqnarray}
\mu_u =\mu_c =\mu_t \equiv \mu  \label{mu1}\\
\mu_d =\mu_s =\mu_b \equiv {\tilde \mu  } \label{mu2}
\end{eqnarray}
However, upper heavier quarks decay into lower lighter ones via the reaction
\begin{equation}
 Q \rightarrow q + W \rightarrow q + {l} + {\bar
\nu}_l\label{eq:decay} \end{equation}
(Top quarks decay almost exclusively via $t \rightarrow b +
W$.)
The lepton conservation is insignificant in our calculation
because the neutrinos have negligible  partial pressure and
they can escape soon after their production. Thus we keep only the
chemical potential for electrons, muons and taus, which must be equal:
$\mu_e=\mu_{\mu} = \mu_{\tau}\equiv \mu_l $. Moreover,  the
decay equation (\ref{eq:decay})
couples the chemical potential of the quark doublets:
\begin{equation}
{\tilde \mu} = \mu + \mu_l \label{mu3}
\end{equation}
 Thus we have only two  variables left.
 If we assume local electric charge neutrality then there will remain only one
independent variable, say, $\mu$, to describe the quark-lepton matter
at zero temperature.

Now let us summarize the necessary expressions for the density, energy density
and pressure of the quark-lepton matter.
For a particle of given type the thermodynamic quantities are the following
(with $\hbar = c = 1$):
 \begin{eqnarray}
n_i &= { g_i \over{(2 \pi)^3 }} \int_{m_i}^{\mu_i}\limits \  d^3
k &=
      { g_i \over{6 \pi^2} } (\mu_i^2 - m_i^2)^{3/2} \\
\varepsilon_i &= { g_i \over{(2 \pi)^3 }}
      \int_{m_i}^{\mu_i}\limits \varepsilon_i (k) \  d^3 k  &=
      { g_i \over{ 8 \pi^2}} \big[ \mu_i
      (\mu_i^2 - m_i^2)^{1/2}
      (\mu_i^2 - {1 \over 2} m_i^2) +
      {m_i^4 \over 2}  \log{m_i \over{\mu_i +
   (\mu_i^2 - m_i^2)^{1/2}} }        \big]  \\
p_i &= n_i {de_i \over{ dn_i}} - e_i &=
      { g_i \over{ 24 \pi^2 }} \big[ \mu_i
      (\mu_i^2 - m_i^2)^{1/2}
      (\mu_i^2 - {5 \over 2} m_i^2) -
      {3 \over 2} m_i^4 \log{m_i \over{\mu_i +
   (\mu_i^2 - m_i^2)^{1/2}}  }       \big]
\end{eqnarray}
Here the one-particle energy is $\varepsilon_i = \sqrt{k^2 + m_i^2}$ ,
$g_i = 6$ is the degeneracy for quarks ($i=u,d,s,c,b,t$)
and $g_i=2$ is the degeneracy for leptons ($i=e^-, \mu^-,\tau^-$).

The total energy and pressure of the quark-lepton matter is the following:
\begin{eqnarray}
\varepsilon  &= \sum_{i} e_i \Theta (\mu_i^2 - m_i^2) +  B^4 \label{eq:e} \\
p &= \sum_{i} p_i \Theta (\mu_i^2 - m_i^2) -  B^4 \label{eq:p}
\end{eqnarray}
$\Theta (x)$ is the Heaviside function, B is the phenomenological
bag constant and the chemical potential can be determined from Eqns.
(\ref{mu1},\ref{mu2},\ref{mu3}) as the function of $\mu$ and
$\mu_l$.

Plugging in Eqs. (\ref{eq:e},\ref{eq:p}) into Eq. (\ref{eq:TOV})
and
using the mass continuity condition (\ref{eq:MassCont}) we still
need another equation to connect our independent variables $\mu$
and $\mu_l$.

The condition of local charge neutrality yields the necessary
expression to determine the quark chemical potential $\mu$
dependence of leptonic chemical potential $\mu_l$:
\begin{equation} \sum_{i} n_i  q_i \Theta (\mu_i^2 - m_i^2) \sgn \mu_i = 0
\label{eq:charge}
\end{equation}
where $q_i = 2/3$ is the electric charge for ($i=u,c,t$),
$q_i=-1/3$ is the charge for ($i=d,s,b$) and $q_i=-1$  for leptons
($i=e^-, \mu^-,\tau^-$).

\subsection{The system of differential equations to  be integrated
           }
\label{sec:SDES}

We have Eqs. (\ref{eq:TOV},\ref{eq:MassCont},\ref{eq:charge})
to be solved numerically in order
to obtain an equilibrium configuration of a dense stellar object.

This system of  equations is
governed by ${\mu}$ of Eq.(\ref{mu1})
which determines the (central) energy density, i.e. the source of
gravitation. The technical details of some numerical problems,
due to the shoulder in the density profile, are discussed in
Appendix \ref{numdiffic}.

The phase transition into hadronic matter occurs where
\begin{eqnarray}
p_H = p_{QGP}
\nonumber \\
{d\varepsilon \over{dn_H}} = \mu_H = 2 \mu_d + \mu_u = 3 \mu_u + 2
\mu_e \end{eqnarray}
We neglect the pressure coming from the leptons at this stage.

{}From this point on we use a hadronic EOS\cite{walecka} with parameters
$c_v^2=195.7$ and $c_s^2=266.9$ until the surface of the stellar
object is reached, i.e. where $p_H = 0$. This $T=0$ EOS exhibits
nuclear saturation and has been
derived from a relativistic mean field theory incorporating neutral
scalar and vector meson fields. The two dimensionless coupling
constants of the theory were matched to the binding energy and density
of infinite nuclear matter in order to obtain results at all densities.

 Since we are discussing cold matter, the equations for the energy
density and pressure are equivalent to knowing $T_{\mu \nu}$.
 This treatment is fully relativistic.

 We checked our model against a simple EOS, quadratic in the nuclear
 density, as well which yielded similar results.

\subsection{The shortcomings of our approach}
\label{sec:shortcomings}

Let us think over what the main deficiencies in our treatment may be.

As for the quark matter, we used a phenomenological bag model with
a bag constant and current quark masses as parameters. The bag constant
$B$, which parametrizes the difference of the exact and perturbative
vacua  (thus actually accounts for interactions among quarks)
is measured only in hadron spectroscopy at low energies, so it is
rather uncertain and this is inherited  by  both $\varepsilon$ an $p$.
Similarly, the EOS was an asymptotically free quark plasma which is
clearly a simplification. However to this point we can note that at
$174$ GeV the quark plasma is expected to be fairly perturbative.

At the $t$ peak we are already at energies well in the order of magnitude
of the Weinberg-Salam mixing of electroweak interactions. While in itself
this mixing may be of secondary importance for the EOS, a possible
accompanying first order phase transition would not be insignificant
since then a latent heat/compression energy of the order of magnitude
$\frac{E_0^4}{(\hbar c)^3}$ would appear simulating a substantial
another "bag constant". It is not clear yet if the Weinberg-Salam
symmetry breaking transition is a first order transition or not. However
no doubt that $m_W = 80$ GeV and $m_Z = 91$ GeV, well below $m_t$.

For hadron matter we used a Walecka-type EOS whose validity range is
$\rho \geq 5 \; 10^{14} g \; cm^{-3}$. Also, it definetely
cannot be relevant below $ \approx 10^{12} g \; cm^{-3}$ where
inverse $\beta$-decay sets in. However, as it turns out, the shell
where our hadronic EOS breaks down carries only some
 hundredths of solar mass.
Anyway, our aim was not to get accurate numbers for masses or radii but to
qualitatively
investigate a central density regime not explored as far as we know.

We neglected electron pressure in the hadronic matter which is
significant at matter densities in the order of $\rho \approx 10^{13} g
cm^{-3}$ and could change our results for the radii.

We shortcutted the subtle question of phase transition by saying
neutrons dominate. However cca. $10$ \% protons (and electrons to
ensure charge neutrality) and probably other hadronic particles, up to
 a few per cents, like $\Delta$'s and heavy mesons also contribute and
they may change the phase structure. For a truly satisfactory discussion
we would need a single EOS  which contains, as limiting cases, both the
hadronic and quark matter EOS.  The less explicite (weaker first order
or possibly second-order) the deconfinement phase transition is, the
less acute this problem of supercompressing and overrarifying is
\cite{kampluk}.

\section{Conclusion}
\label{sec:results}

Our results are summarized in a series of figures. The masses and
radii of
quark stars have been calculated at cca. $340$ different central
chemical
potential values, and even then numerical techniques were needed to smoothen
the curves since at high central densities the inhomogeneities are substantial
to cause numerical difficulties.

\begin{figure}
          \begin{center}
          \leavevmode\epsfysize=3.0in
          \epsfbox{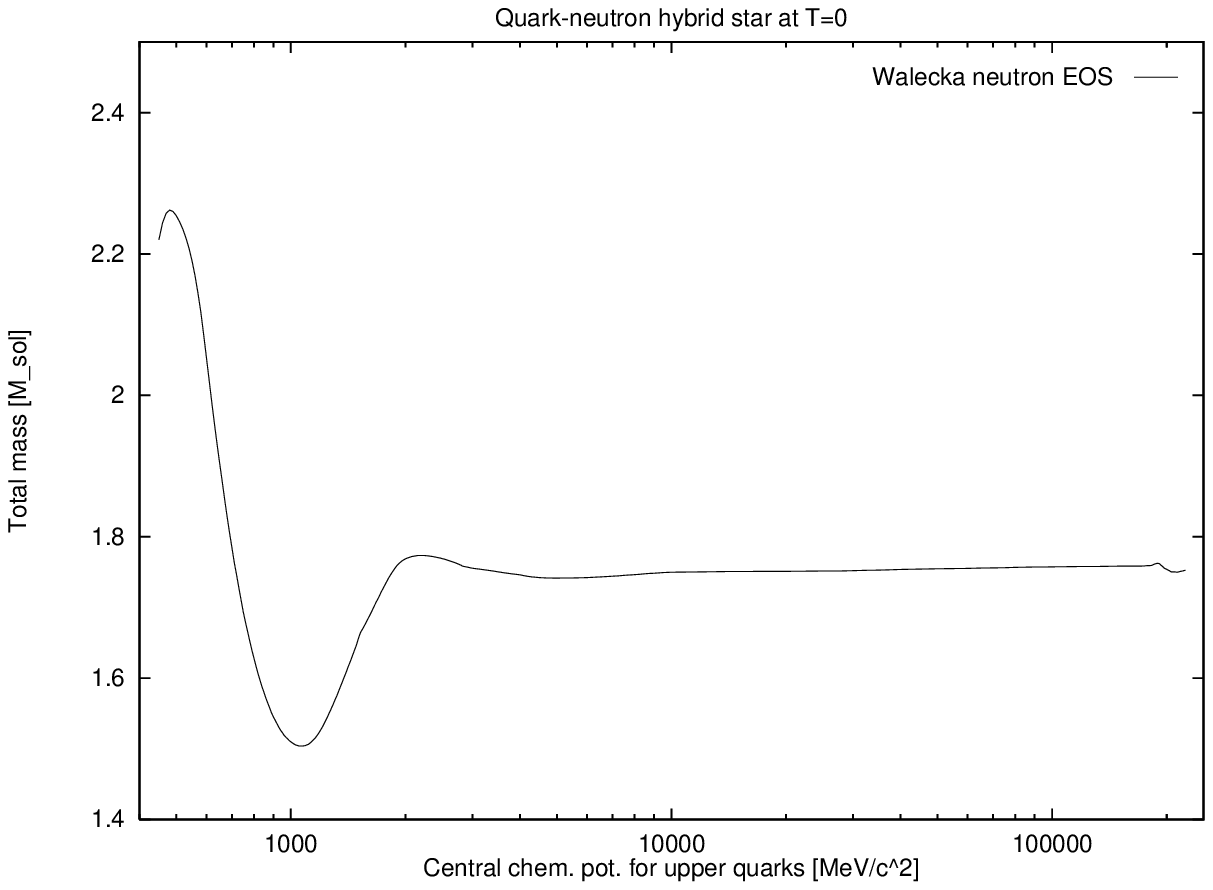}
          \end{center}
\caption{ Mass dependence of hybrid stars on the central chemical
potential $\mu$.
}
\label{fig:mass}

          \begin{center}
          \leavevmode\epsfysize=3.0in
          \epsfbox{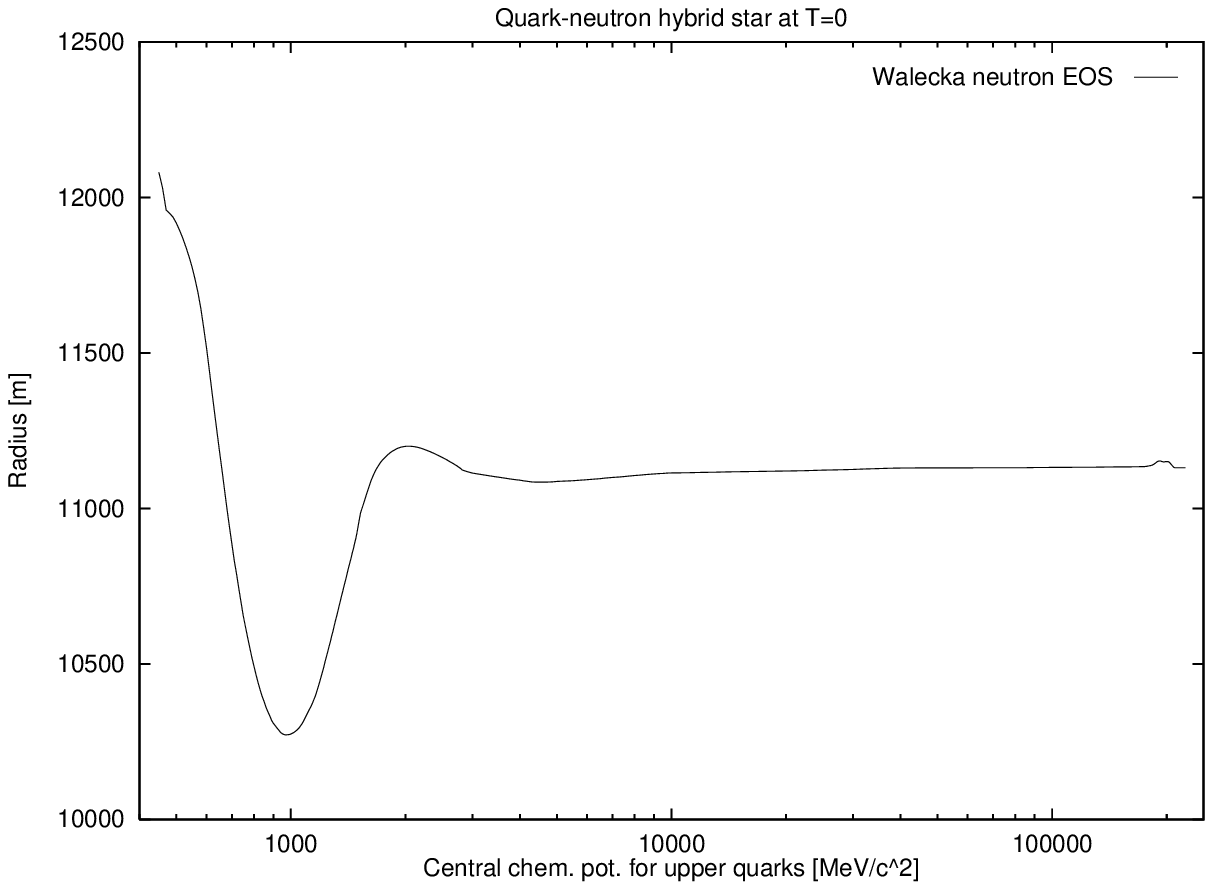}
          \end{center}
\caption{ Radius dependence of hybrid stars on the central chemical
potential $\mu$.
}
\label{fig:radius}
\end{figure}

Fig. \ref{fig:mass} displays the total mass of the equilibrium configurations
vs. central chemical potential (monotonous with central energy density). The
curve starts at the point where the quark plasma appears in the center.

Fig. \ref{fig:radius} is the same but for radius. The curve is qualitatively
similar to Fig. \ref{fig:mass}.

\begin{figure}
          \begin{center}
          \leavevmode\epsfysize=3.0in
          \epsfbox{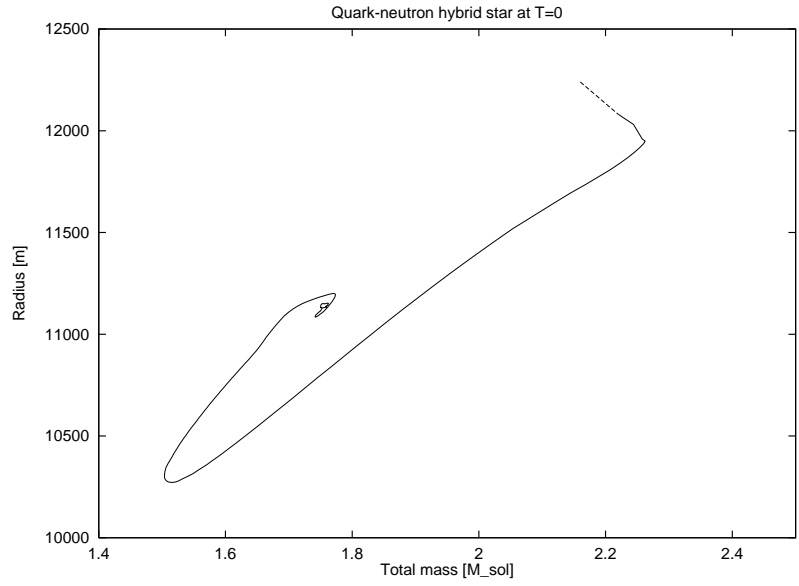}
          \end{center}
\caption{Total mass vs. radius of hybrid stars.
}
\label{fig:massrad}
\end{figure}

Fig. \ref{fig:massrad} composes the two earlier figures and depicts
$R(M)$.   By means of this curve one can directly apply the stability
criteria of Section \ref{sec:stabil}. For increasing central density
the curve starts in the upper right corner, where quark matter appears,
 and ends up in the middle, where top Fermi gas could exist in the very
inside of a hybrid star.

Note that prior to this point the star is an ordinary neutron star with
Walecka-type equation of state. (Some configurations are shown with
dashed line.)  There $M(\varepsilon_0)$ is on the rise and this upward
slope was found to be stable. (see e.g. Ref. \cite{harr}). This turning point
corresponds to the Landau-Oppenheimer-Volkoff point\cite{landau}

At the first turning point (point 1) on Fig. \ref{fig:massrad} $R$ is
growing at a maximal $M$. Consequently the mode of lowest frequency
becomes unstable. The next extremal point is a mirror turn (turn to the
right on the left side), therefore the lowest frequency stable mode
becomes unstable again. This point corresponds to the Misner-Zapolsky
point\cite{misner2}.

The form of the quark equation of state  is qualitatively similar to
the EOS  of the Fermi gas of non-interacting neutrons of the early
calculations and this may explain the qualitatively similar structure
of our $R(M)$ curve to that of Ref. \cite{harr}.

Beyond this point $M$ is again increasing; we see the asymptotically
damped oscillation of $M(\varepsilon_0)$ (or $M(\mu)$) as demonstrated
in Ref. \cite{harr}.

However not too far from the M-Z point $c$ quarks appear in the core,
later they become relativistic and, as usual, the $M(\varepsilon_0)$
($M(\mu)$) curve stops to increase while reaching the relativistic
Fermi regime. This fact puts point 3 to slightly beyond chemical
potential $2$ GeV.  Hence another stable mode becomes unstable, similarly
as found earlier for pure non-interacting neutron matter.

Fig. \ref{fig:topnagy} shows that we lose still another stable mode at
the next turning point (point 4). Comparison to Fig. \ref{fig:mass}
gives that this turning point is located at $\mu = 5000$ GeV, which
corresponds to the appearance of $b$ quarks beyond $4500$ GeV. The minimum
is extremely shallow and wide.

The next turning point (point 5), located at $192$ GeV, restores one
mode as for stability. The peak is very small ($0.01$ $M_{\odot}$ compared
to the background) but rather expressed; it seems to belong to $t$ quarks.

Since no new degrees of freedom may appear in the equation of state beyond
$t$ quarks, we stopped the calculation after this turning point. Thenceforth
the $M(\varepsilon_0)$ curve must display the well-known asymptotically
damped oscillation with icreasing instability.

\begin{figure}
          \begin{center}
          \leavevmode\epsfysize=3.0in
          \epsfbox{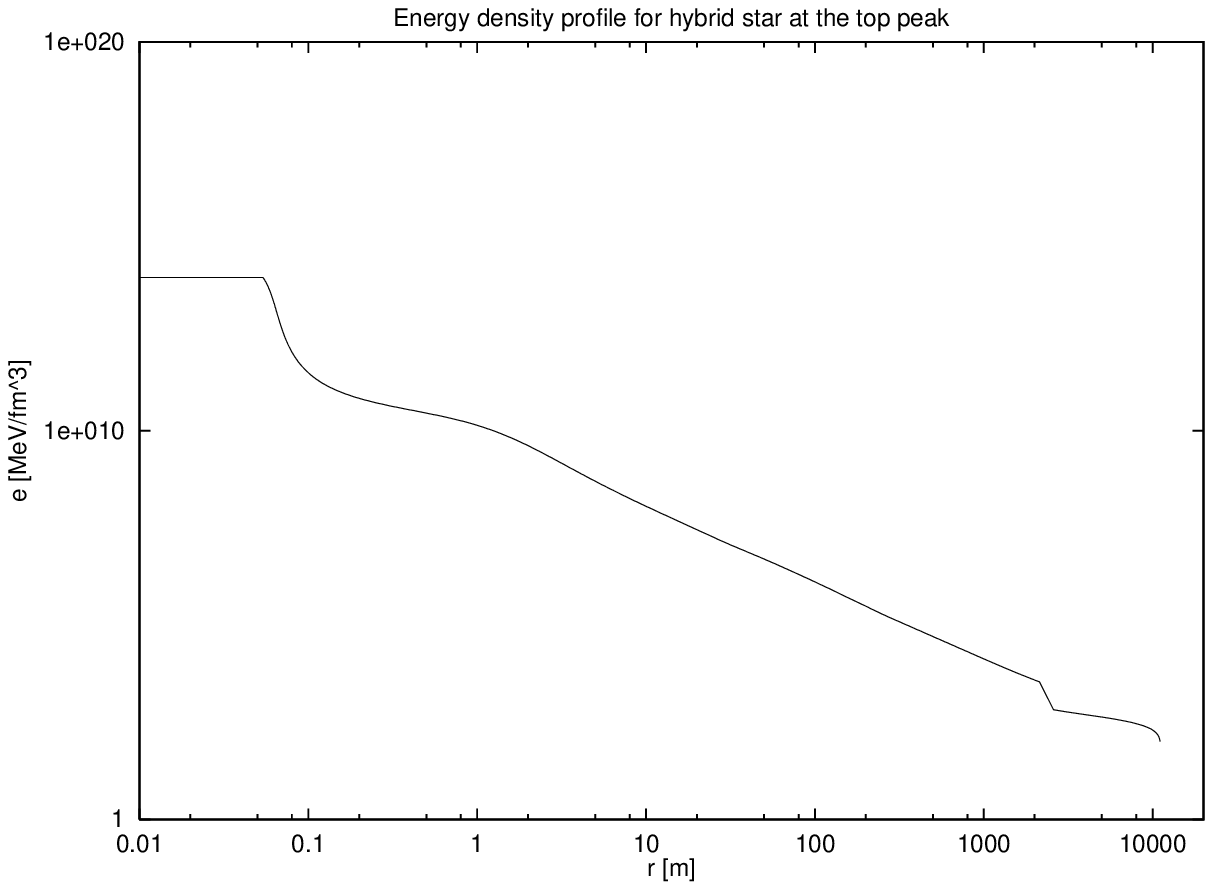}
          \end{center}
\caption{ Energy density vs. radius in the model hybrid star with central
density at the the top peak.  } \label{fig:energyprofile}

          \begin{center}
          \leavevmode\epsfysize=3.0in
          \epsfbox{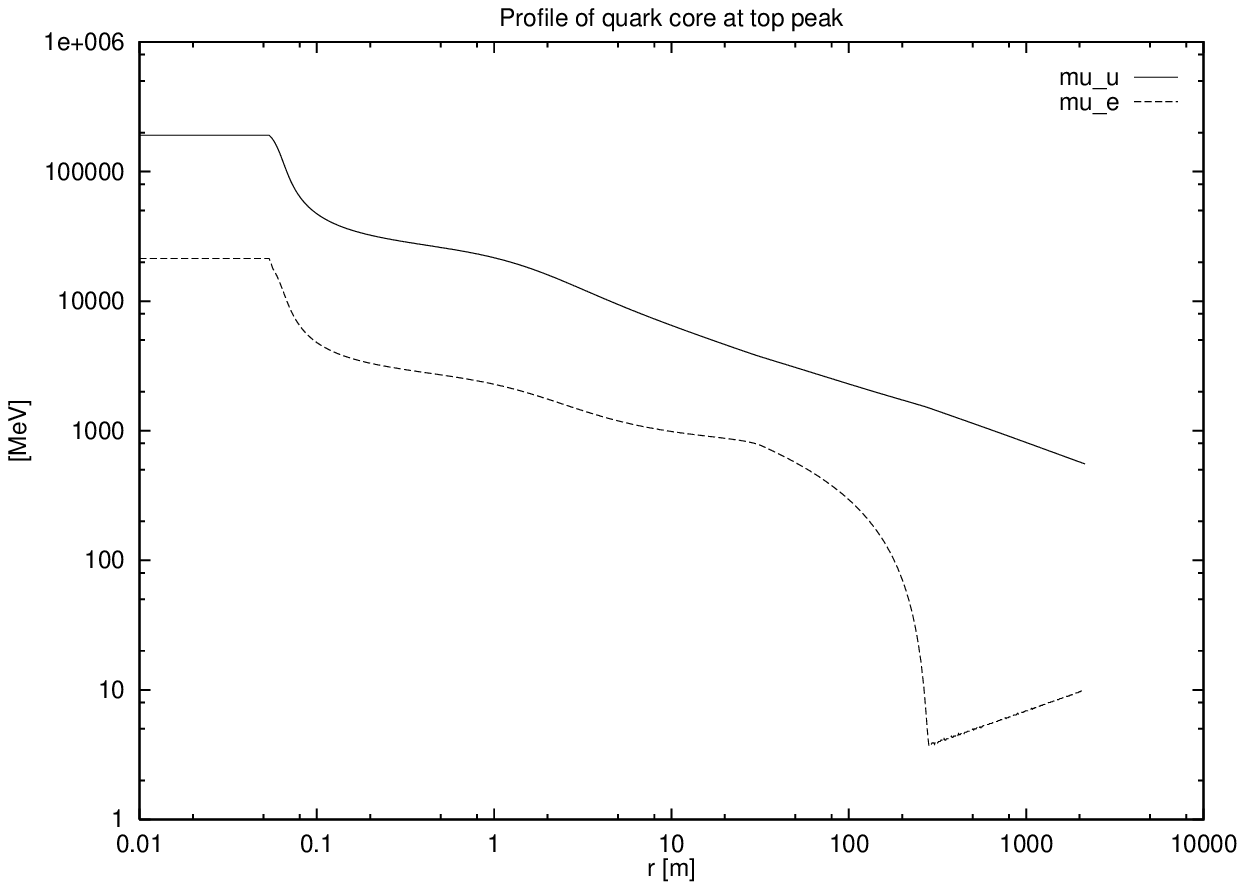}
          \end{center}
\caption{ Profile of the model hybrid star with central density
corresponding to the top peak.
}
\label{fig:profil}
\end{figure}

Fig. \ref{fig:energyprofile} shows the energy density profile at the
$t$ peak. Note the sudden drop in the $10$ cm range which coincides
with the disappearance of $t$ quarks. The jump of the first order
phase transition is at cca. $2$ km.

Fig. \ref{fig:profil} is the change of the two chemical potentials,
as connected through charge neutrality, in a configuration at the $t$ peak.
The $\mu$ curve decreases smoothly and monotously but $\mu_l$ shows a
turning point at about $300$ m. This surprising behaviour needs an
explanation.

The $u$, $d$, $c$,$s$,$e^-$ and $\mu^-$ number densities are shown
in Fig. \ref{fig:partdens}. (In this range $0 \leq \mu \leq 2000$ MeV,
$n_{\tau} = 0$.) Now observe that descending with $\mu$ the electron
density must drop sunstantially at $1500$ MeV since there
the $c$ quarks  vanish and there remain two negative light quark flavours
with $-1/3 \; e$ charge each to compensate for the single positively
charged one with $+2/3 \; e$. However $n_{e^-}$ must not drop to $0$
since $n_s$ is slightly below $n_u$ or $n_d$ due to the higher $s$ mass.
Much later, at $\mu_u + \mu_e \approx 150$ MeV the $s$ quarks vanish
and then a great amount of electrons will be needed to balance the double
charge of $u$ over $d$. This way $\mu_{e^-}$ must slowly increase with
decreasing $\mu_u$ after a minimum which happens to occur at $\mu = 1420$
MeV.

Ref. \cite{gled} calculated quark stars up to charm, but without assuming local
neutrality and  a hadronic mantle. They did find a separate $s$ peak,
but no peak
for light quarks. However the assumptions of the two models were sufficiently
different to give such different results.

We may conclude that, at least in the present approximation, no stable
configuration has been found beyond the LOV point where the star
is mainly a neutron star but with $uds$ plasma in the centre.  If our
approximations are valid then no $c$, $b$ or $t$ quarks are expected to
exist in the present Universe at least under natural conditions.

The present paper is only a first approximation to the problem.
We listed the serious simplifications in Section \ref{sec:shortcomings}.
 Some points will be rediscussed  in a future work. Note that e.g. at the
 starting point of Figs. \ref{fig:mass} -- \ref{fig:topnagy} these calculations
 would need nonperturbative QCD results, practically unavailable now.

\acknowledgements
 This work was supported by the Hungarian Science Fund, OTKA No. T014213.

\appendix

\section{How to avoid the numerical difficulties at $r=0$ }

Since Eq. (\ref{eq:TOV}) is singular at $r=0$ one has to start
integrating at a small positive $\bar{r}$. At high densities it is
necessary to take into account the mass inside this sphere of
radius $\bar{r}$. Therefore we applied a first-order
expansion of Eq. (\ref{eq:TOV}). Then one can choose the starting
point of the numerical integration at
 $r=\bar{r}$ where the initial (central) density (or the equivalent
 $\mu$ ) drops to, say, 95 percents.
This $\bar{r}$ can be expressed as
$\bar{r} = \sqrt{ 0.05 \mu^4/C} $ where
$C = 2 \pi G ( \varepsilon (r=0) + p (r=0))
             ( \varepsilon (r=0)  /3  + p (r=0))$.

\section{How to avoid the numerical difficulties with the sharply
 decreasing density profile}
\label{numdiffic}

Since at densities where top quarks appear one has numerical
difficulties with the integration of Eqs. (\ref{eq:TOV},
\ref{eq:MassCont}, \ref{eq:charge}, \ref{eq:e}, \ref{eq:p} ),
we resorted to various expansions.

At infinite central energy density $\rho = K r^{-2}$ is a solution to
Eq. (\ref{eq:TOV}). One gets the value of $K={{3 c^2} \over
{ 56 \pi G}}$ by comparing the powers in $r$.

We get a regular solution at $r=0$ if we require that
$\rho = {K\over{r^2}} f(r)$ is a regular function where $f(r)$ is
an unknown function we find by substituting into Eq. (\ref{eq:TOV}).
This way the density profile around $r=0$ is
\begin{equation}
\rho (r) = \rho_0 - { { 32 \pi G \rho_0^2} \over{3 c^2}} r^2 +
              { { 7936  \pi^2  G^2  \rho_0^3} \over{3 c^4}} r^4 +
\dots \label{eq:expansion1}
\end{equation}
We used an expansion up to $16^{th}$ order that breaks down, at
densities
relevant to the study of top matter, within a few centimeters.
{}From that radius we applied another expansion as follows.

Performing perturbation calculation
 around the $\rho = K/r^2$ solution
 ( i.e. now we seek a
solution of the form ${K\over{r^2}}(1+\epsilon (r))$ where $\epsilon$ is
small )
which belongs to the infinite central density,  one obtains an
expansion at
$r=r_0$ where $r_0$ is the endpoint of the validity range of
Eq.(\ref{eq:expansion1}):
\begin{equation}
\rho (r) = {K \over{r^2}}{  (1 + a_1 r + 0.464286  a_1^2 r^2 + \dots)}
\label{eq:expansion2} \end{equation}
Here parameter $a_1$  remains free to allow to join expansion
(\ref{eq:expansion1}) and expansion (\ref{eq:expansion2}) in a
continous way.

We designated an arbitrary density (say, which belongs to $\mu = 25
$  GeV,
 so that the relation $P \approx {1 \over{3 c^2} \rho}$ still holds )
at which we continued the numerical solution of the set of equations.

\vfill\eject

\vskip3.2in
\begin{figure}
\caption{Total mass vs. radius in the vicinity of the top peak.
}
\label{fig:topnagy}
\end{figure}

\eject\vfill
\vfill

\pagebreak
\vfill
\vskip6.5in
\begin{figure}
\caption{Particle number densities vs. chemical potential $\mu$.
}
\label{fig:partdens}
\end{figure}
\end{document}